\documentclass[%
	aip,
	jap,
	reprint,%
	amsmath,%
	amssymb,%
	superscriptaddress,%
	floatfix,%
	twocolumn,%
]{revtex4-1}

\usepackage{graphicx}
\usepackage{epstopdf}
\usepackage{dcolumn}
\usepackage{bm}

\renewcommand\vec{\mathbf}

\usepackage[%
	colorlinks=true,%
	citecolor=blue%
]{hyperref}

\usepackage[utf8]{inputenc}
\usepackage[T1]{fontenc}
\usepackage{mathptmx}

\begin{document}

\title{Raman spectroscopy of alpha-FeOOH (goethite) near antiferromagnetic to paramagnetic phase transition\texorpdfstring{\vspace{2ex}}{}}
\author{M. V. Abrashev}
\affiliation{Sofia University, Faculty of Physics, 5 James Bourchier Blvd., 1164 Sofia, Bulgaria}
\author{V. G. Ivanov}
\affiliation{Sofia University, Faculty of Physics, 5 James Bourchier Blvd., 1164 Sofia, Bulgaria}
\author{B. S. Stefanov}
\affiliation{Sofia University, Faculty of Physics, 5 James Bourchier Blvd., 1164 Sofia, Bulgaria}
\author{N. D. Todorov}
\email[]{neno@phys.uni-sofia.bg}
\affiliation{Sofia University, Faculty of Physics, 5 James Bourchier Blvd., 1164 Sofia, Bulgaria}
\author{J. Rosell}
\affiliation{\hbox{Rosell Minerals, C. Industries 57, entr. 3a., 08820 El Prat de Llobregat, Barcelona, Spain}}
\author{V. Skumryev\vspace{1.5ex}}
\affiliation{Departament de F\'{i}sica, Universitat Autònoma de Barcelona, 08193 Barcelona, Spain}
\affiliation{Instituci\'{o} Catalana de Recerca i Estudis Avancats, 08010 Barcelona, Spain}

\date{\today}

\begin{abstract}
Synthetic powder, ore samples and mineral single crystals of goethite ($\alpha-$FeOOH) were investigated with polarized Raman spectroscopy at temperatures from $300$~K to $473$~K. The symmetry of the vibrational modes, observed in different scattering configurations, was determined unequivocally. The assignment of the Raman-active modes to definite atomic vibrations is supported by two types of lattice-dynamical calculations: empirical shell-model and $\textit{ab initio}$ DFT. The temperature dependencies of the lineshape parameters of some Raman-active vibrations near to the antiferromagnetic-paramagnetic phase transition infers for a significant spin-lattice coupling in this compound. The most informative in this aspect is the $B_{3g}$ phonon at $387$~cm$^{-1}$, which overlays a broad scattering background and displays a pronounced asymmetric Fano-lineshape. The asymmetry increases in the paramagnetic state above the Neel temperature ($T_{\text{N}} = 393$~K) indicating a strong interaction of this mode with the underlying excitation continuum. The origin of the excitation background is discussed in light of our experimental results and the existing data for the magnetic structure and transport properties of $\alpha-$FeOOH. We rationalize that, most probably, the background stems from magnetic excitations, and the asymmetric shape of the $B_{3g}$ phonon is a result of a linear spin-phonon coupling of this mode with the Fe-O1-Fe spin dimers. Another mechanism, a phonon interaction with thermally activated charge carriers above above $T_{\text{N}}$, is also considered.  
\end{abstract}
\maketitle

\section{Introduction}
The transition metal oxides are a large group of materials exhibiting a variety of structural, electrical and magnetic properties. Among them the iron oxides are most popular due to their largest abundance as ores and minerals in Earth crust.\citep{iron_oxides} In the family of FeOOH hydroxide polymorphs $\alpha$-FeOOH (goethite) is the compound, which is the most thermodynamically stable at ambient conditions. At room temperature goethite is an antiferromagnetic (AF) insulator and exhibits transition to a paramagnetic (PM) state at a Neel temperature of $T_{\text{N}}\approx 400$~K. Systematic studies of its properties in the paramagnetic state are limited because on heating at temperatures above $600$~K it decomposes to $\alpha$-Fe$_{2}$O$_{3}$ (hematite) and water.\citep{de_Faria_2007} Moreover, at equilibrium conditions the temperature of the goethite/(hematite + water) transformation is even lower (about $400$~K).\citep{iron_oxides}

Goethite is one of the most common antiferromagnetic materials in nature. Nevertheless, its intrinsic physical properties are not fully understood yet. The reason is that the goethite samples, both natural and synthetic, are usually composed of nano- or micronsize crystallites often with poor crystallinity, different degree of preferential orientation, exact stoichiometry and impurities.
Recently, the interest to $\alpha$-FeOOH has been attracted by the theoretical prediction of a significant linear magnetoelectric effect in the AF state.\citep{Ter_Oganessian_2017}
If this prediction is approved, the goethite will be established as one of the most prominent high temperature magnetoelectric materials.
In this aspect Raman spectroscopy is a powerful tool for investigation of fine changes in the crystal structure, the electrical, and the magnetic properties of materials, through their impact on the $\Gamma$-point optical phonons. For example, perovskite manganese oxides display significant frequency and linewidth renormalization of their Raman-active modes at temperatures near the AF phase transition,\cite{Granado1998,Granado1999} which give a wealth of information about the mechanism of spin-lattice interaction in this class of materials.\cite{Laverdiere2006,Flores2006} So far, characteristic Raman frequencies of goethite have been documented widely in the literature but Raman spectroscopy has been used mainly for identification of this compound in multiphase mineral samples.\cite{Kustova_1992, de_Faria_1997, Oh_1998, LEGODI_2007, de_Faria_2007, Hanesch_2009, Nieuwoudt_2010, Kreissl_2016, Hedenstedt_2017, Liu_2019} Therefore, the main goal of the present work is to assess the importance of spin-lattice interaction in $\alpha$-FeOOH by investigating: (i) the symmetry of the Raman-active modes observed in the Raman spectra; (ii) their assignment to specific atomic displacements, and (iii) the temperature behaviour of the Raman lines in the vicinity of the AF-PM phase transition. To our best knowledge, these issues have not been addressed in the literature, up to now.

In this paper at first we determined the symmetry of the lines observed in the polarized Raman spectra of goethite ores, using the fact that due to the needle-like shape of the goethite microcrystals, even when their sizes are submicronic, their orientation is correlated. After that we precise their assignment using sufficiently large for micro-Raman spectroscopy mineral single crystals with determined orientation. The experimental findings were compared with lattice-dynamical calculations. As a results $22$ out of the $24$ Raman-active modes were assigned to observed lines in the Raman spectra. Later, both unpolarized and polarized Raman spectra were obtained in the temperature interval $293$~K -- $473$~K. After fit of the observed in the spectra lines it was found that some of the lines show anomalies in the temperature dependencies of their parameters (position, width and intensity). Moreover, one of them, the $B_{3g}(3)$ line at $387$~cm$^{-1}$ shows asymmetric shape, which asymmetry increases when the temperature increases towards the $T_{\text{N}}$ and remains almost constant in the PM state. The line profile can be fitted with Fano-shape. The origin of this line behavior is discussed.

\section{Experimental methods and calculation details}
\begin{figure}[htbp]
\centering
\includegraphics[width=\linewidth]{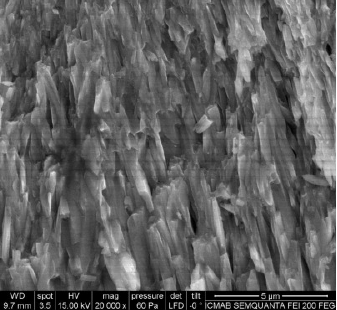}
\caption{\label{SEM1}Electron microscope image of surface of goethite ore}
\end{figure}

\begin{figure}[htbp]
\centering
\includegraphics[width=\linewidth]{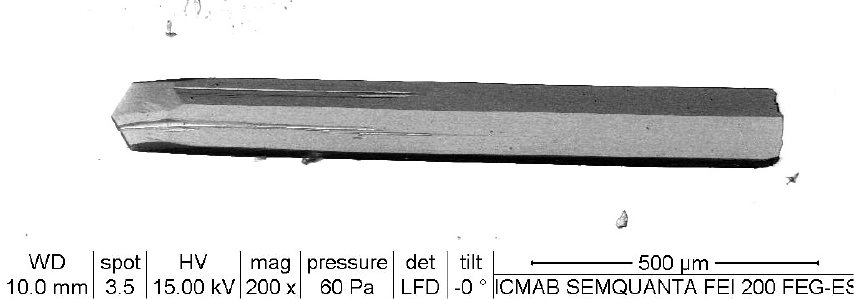}
\caption{\label{SEM2}Electron microscope image of goethite single crystal}
\end{figure}

The origin of the goethite ores is mine Kremikovtsi (Bulgaria). The source of the mineral needle-like single crystals with length up to $2$~mm and diameter up to $100~\mu$m is Tounfit region (Morocco). Both ores and single crystals were characterized using scanning electron microscopy (see Figs.~\ref{SEM1},~\ref{SEM2}). The EDX analysis show no presence of other chemical elements except Fe and O.

\begin{figure}[htbp]
\centering
\includegraphics[width=\linewidth]{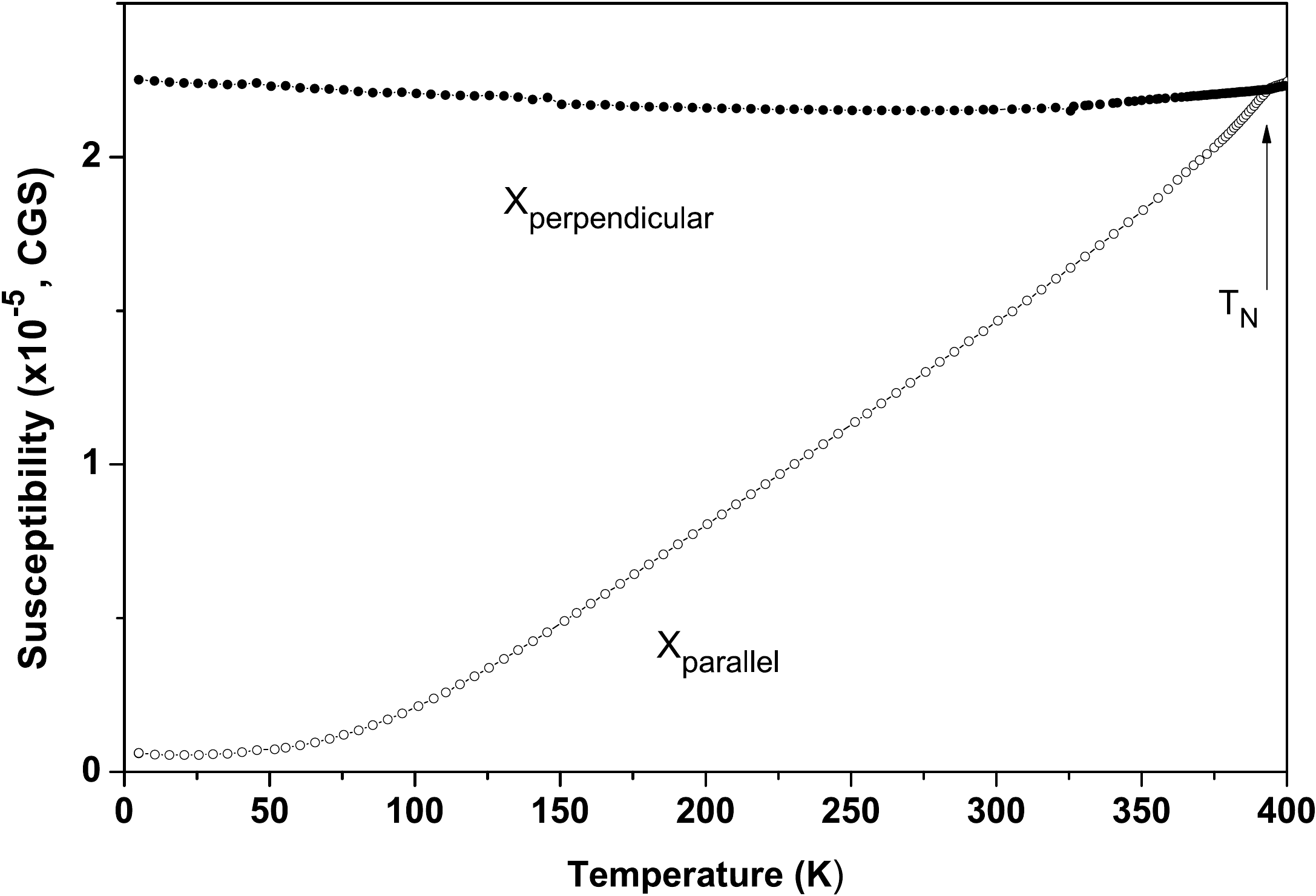}
\caption{\label{magnetic}Temperature dependence of the DC magnetic susceptibility measured in field of $4$~T applied along the $[010]$ crystallographic direction (the AFM axis) and perpendicular to it.}
\end{figure}

The magnetic response was measured by SQUID magnetometer on a sample composed of several crystals with their $b$-axes aligned. The temperature dependence of the DC magnetic susceptibility measured in field applied along the $b$-axis as well as in direction perpendicular to it is reminiscent of one for collinear antiferromagnet (see Fig.~\ref{magnetic}). With the increase of the temperature, a magnetic phase transition from antiferromagnetic to paramagnetic phase is observed with Neel temperature, $T_{\text{N}} = 393$~K. We note, that $T_{\text{N}}$ for the other two types of samples used in this study---the synthetic powder and the goethite ores is $393$~K and about $347$~K, respectively.

Raman spectra were obtained using LabRAM HR Visible (HORIBA Jobin Yvon) Raman spectrometer in backscattering configuration. For excitation were used several laser lines ($633$~nm of He-Ne laser and $515$~nm and $458$~nm of Ar$^{+}$ laser). For the spectra at room temperature an $\times 100$ objective and $600$~mm$^{-1}$ grating were used (the spectral distance between CCD pixels is $1.0$~cm$^{-1}$). For collection of spectra at different temperatures LINKAM THMS600 heating cell was used in combination with a $\times 50$ objective with long working distance and $1800$~mm$^{-1}$ grating (the spectral distance between CCD pixels is $0.3$~cm$^{-1}$). The direction of the polarization of the incident linearly polarized laser light was changed using a $\lambda/2$ plate. The scattered light was analyzed with a polarizer. The different scattering configurations were realized rotating the sample and/or microscope table in the laboratory coordinate system. The optimal laser power on the laser spot (with diameter about $2~\mu$m) on the sample surface (determined after power tests to ensure that there is no local laser overheating) was $0.07$~mW for the powder, $0.4$~mW for both the ore samples and single crystals when $\times 100$ objective was used (for room temperature measurements), and $1.0$~mW for temperature measurements of the single crystals (when $\times 50$ objective was used). The procedure of the power test consists of comparison of the line parameters (position and linewidth) of series of Raman spectra, obtained with increasing laser power and decreasing accumulation time, so that the product of laser power and accumulation time is constant for all spectra. When an overheating occurs, the position of the lines shifts to lower frequencies and their linewidth increases. 

Lattice-dynamical calculations (LDC) in the framework of the shell model\cite{Dick1958,Woods1960} were performed using the General Utility Lattice Program.\cite{Gale1997,Gale2003} In this model an ion consists of a rigid \textit{shell} with charge $Y$, which represents the valence electrons and has no mass, whereas the nucleus and the inner electrons form the \textit{core}, which has all the ion's mass. The core is bound to the shell by harmonic restoring forces of spring constant $k$, thus the ion polarizability can be introduced as $Y^{2}/k$. The short-range interactions between non-bonded atoms were modeled with repulsive Born-Mayer potential in the Buckingham form $\displaystyle U(r) = A \exp{(-r/\rho)-C/r^{6}}$. In addition, the covalent O-H bond was described by the Morse potential, $\displaystyle U(r)=D_{e}\left\{1-\exp{\left[-a(r-r_{0})\right]}\right\}^{2}$, where $r_{0}= 0.9485$~\AA\ is the equilibrium O-H bond length. All other parameters $Y$, $Z$, $k$, $A$, $\rho$, $C$, $D_{e}$, and $a$ are listed in Table~\ref{tab:LDCsm}. The cell parameters and atomic positions used in the calculations, are taken from Ref.~\onlinecite{Yang_2006}. The shell model parameters $k$ and $a$ were refined to achieve the best fit to the experimental data for the lattice structure and the hydrogen vibrations near $1000$ and $3000$~cm$^{-1}$.
\begin{table*}[htbp]
\caption{\label{tab:LDCsm}List of the initial values for the shell model parameters, which were taken from Ref.~\onlinecite{Lewis1985} as well as the provided, with the GULP code, libraries. After the refinement we obtained the following values for $k$ and $a$, $28.86~e^2\!/\!\text{\AA}^3$ and $1.8079~\text{\AA$^{-1}$}$, respectively.}
    \begin{ruledtabular}
        \begin{tabular}{cdddcdddddc}
            Ion                                      &
            \multicolumn{1}{c}{$Z\ (|e|)$}           &
            \multicolumn{1}{c}{$Y\ (|e|)$}           &
            \multicolumn{1}{c}{$k\ (e^2\!/\!\text{\AA}^3)$}&
            Ionic pair                               &
            \multicolumn{1}{c}{$A\ (e\text{V})$}     &
            \multicolumn{1}{c}{$\rho\ (\text{\AA})$} &
            \multicolumn{1}{c}{$C\ (e\text{V\AA}^6)$}&
            \multicolumn{1}{c}{$D_{e}\ (e\text{V})$} &
            \multicolumn{1}{c}{$a\ (\text{\AA$^{-1}$})$} &
            \multicolumn{1}{c}{cutoffs (\AA)}\\
            &&&&&&&&&& min -- max \\\hline
            Fe &  3.00000 &          &       & Fe-O1 &  1102.40 & 0.3299 &  0.000 &        &        & 0.0 -- 12.0 \\
            O1 &  0.86902 & -2.86902 & 74.92 & Fe-O2 &   862.08 & 0.3299 &  0.000 &        &        & 0.0 -- 12.0 \\
            O2 & -1.42600 &          &       & O-O   & 22764.00 & 0.1490 & 27.879 &        &        & 0.0 -- 12.0 \\
            H  &  0.42600 &          &       & O1-H  &   208.11 & 0.2500 &  0.000 &        &        & 0.0 -- 10.0 \\
               &          &          &       & O2-H  &   311.97 & 0.2500 &  0.000 & 7.0525 & 2.1986 & 1.4 -- 10.0 \\
        \end{tabular}
     \end{ruledtabular}
\end{table*}

As a crosscheck of the results obtained by the empirical shell-model, first-principle density-functional theory (DFT) calculations of the phonons were performed by means of Quantum Espresso (QE)\cite{QE} plane wave (PW) program suit. The choice of exchange-correlation functional and atomic pseudopotentials is essential for the accuracy of the DFT calculations on the transition-metal oxydes due to the significant correlations between $d$-electrons of the transition-metal ions. Hubbard-corrected schemes like LDA+$U$ or GGA+$U$ have been shown to improve significantly the convergency of the self consistent field (SCF) calculations and provide a good description of the structural, electronic and magnetic properties of this class of compunds. So far, the reported up to now first-principle calculations on $\alpha-$FeOOH have been performed within GGA+$U$ approximation.\cite{Blanchard_2013,Ter_Oganessian_2017} In QE software DFT+$U$ functionals are implemented for SCF and optimization but not for phonon calculations, which forced us to look for alternative approaches to the lattice dynamics of $\alpha-$FeOOH. Therefore, we made use of the recently proposed Optimized Norm-Conserving Vanderbilt (ONCV) pseudopotentials\cite{Hamann2013} for PBE functional.\cite{PBE1996} For a moderate or no excess of the PW kinetic energy cutoff, compared to the ultrasoft pseudopotentials, the ONCV potentials show an excellent correspondence with the structural data for a variety of materials, including many compounds of transition-metals.\cite{Schlipf2015} The calculations were made on a $4 \times 12 \times 8$ Monkhorst-Pack (MP) $k$-point grid for a PW kinetic energy cutoff of 80~Ry ($\approx 1100$~eV).  Optimized lattice constants of $a =9.87,~b = 3.05$, and $c = 4.50$~{\AA} (in $Pnma$ notation) were obtained, which are in good agreement with the experimental lattice parameters of $\alpha-$FeOOH.  The phonon frequencies, calculated for the relaxed structure, are listed in Table~\ref{tab1}. 

\section{\label{sec:3}Results and discussion}
\begin{figure}[htbp]
\centering
\includegraphics[width=\linewidth]{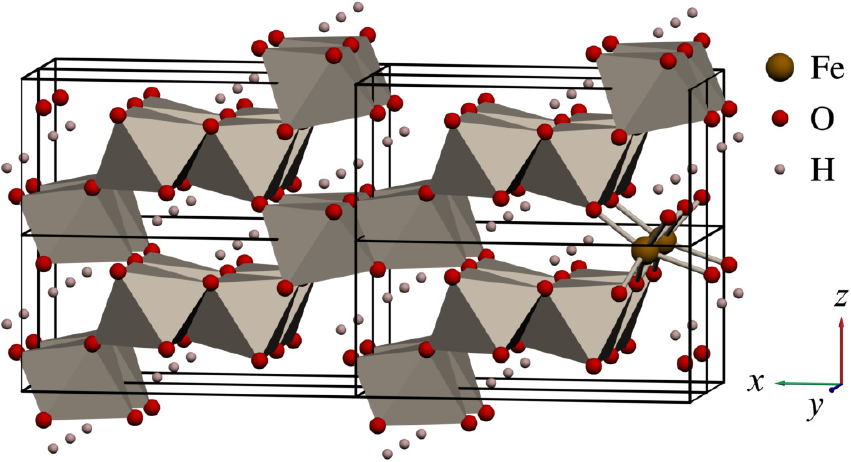}
\caption{\label{structure1} Crystal structure of $\alpha$-FeOOH (goethite). A supercell of $2 \times 2 \times 2$ unit cells is drawn.}
\end{figure}

The crystal structure of $\alpha$-FeOOH is orthorhombic with space group $Pnma$ ($D_{2h}^{16}$, No.~$62$, $Z = 4$, see Fig.~\ref{structure1}).\cite{Yang_2006} All four types of atoms, Fe, O1, O2, and H, occupy positions with same site symmetry (Wyckoff position $4c$). Fe atoms are connected with 3O1 and 3O2 atoms, forming Fe(O1)$_{3}$(O2)$_{3}$ octahedra. Two adjacent along $[010]$ direction FeO$_{6}$ octahedra have common edge, forming chains along $[010]$ direction. The octahedra from two adjacent chains also have common edge, making strongly bonded double chains along $[010]$ direction, leading to the needle-like shape of microcrystals. The adjacent double chains have common oxygen atom (they are corner-shared). The hydrogen atom is bonded to oxygen O2. Each atom from the unit cell participates in normal vibrational modes with irreducible representations $2A_{g} + A_{u} + B_{1g} + 2B_{1u} + 2B_{2g} + B_{2u} + B_{3g} + 2B_{3u}$.\cite{Rousseau1981,Kroumova2003} Among them only $A_{g}$, $B_{1g}$, $B_{2g}$, and $B_{3g}$ are Raman-active. The site symmetry restricts the possible directions of atomic vibrations in the different modes. $A_{g}$ and $B_{2g}$ modes are vibrations in $(010)$ plane, whereas $B_{1g}$ and $B_{3g}$ are vibrations along $[010]$ crystallographic direction. Therefore $24$ lines ($8A_{g} + 4B_{1g} + 8B_{2g} + 4B_{3g}$) originating from one-phonon scattering are expected in the Raman spectra. In the simplest approximation of non-resonant Raman scattering their intensity depends only on the directions of the polarization of the incident ($\vec{e}_{\text{i}}$) and scattered ($\vec{e}_{\text{s}}$) light ($\vec{e}_{\text{i}}$ and $\vec{e}_{\text{s}}$ are the unit vectors along these directions): $I_{\vec{e}_{\text{i}} \vec{e}_{\text{s}}} \propto \left(\vec{e}_{\text{i}}\cdot\hat{R}\cdot\vec{e}_{\text{s}}\right)^{2}$. $\hat{R}$ is the Raman tensor. It is symmetric and for the different types of Raman-active modes its non-zero components (in coordinate system connected with the crystallographic axes) are as follows: $\alpha_{xx} \neq \alpha_{yy} \neq \alpha_{zz}$ (for $A_{g}$), $\alpha_{xy}$ (for $B_{1g}$), $\alpha_{xz}$ (for $B_{2g}$), and $\alpha_{yz}$ (for $B_{3g}$).\cite{Rousseau1981,Kroumova2003}

From simple atomic mass considerations $18$ out of the $24$ Raman-active modes, including mainly oxygen and iron atoms vibrations, should have frequencies below $800$~cm$^{-1}$, whereas six of them, being purely hydrogen vibrations, must be situated near $1000$~cm$^{-1}$ (for bending vibrations) and above $3000$~cm$^{-1}$ (for stretching vibrations). As the direct interaction between hydrogen atoms is weak, their six Raman-active modes are expected to be distributed into three Davydov pairs (pair of modes with identical direction of hydrogen atoms vibrations as the only difference between them is the relative phase of the vibrations of the adjacent hydrogen atoms---in-phase or out-of-phase) with very close or coinciding frequencies. These three pairs must be the stretching O-H vibration pair $A_{g} + B_{2g}$, the bending (in $(010)$ plane) O-H vibration pair $A_{g} + B_{2g}$, and the bending (along $[010]$ direction) O-H vibration pair $B_{1g} + B_{3g}$. The lattice-dynamical calculations as well the observed Raman spectra confirmed these expectations.

\begin{figure}[htbp]
\centering
\includegraphics[width=\linewidth]{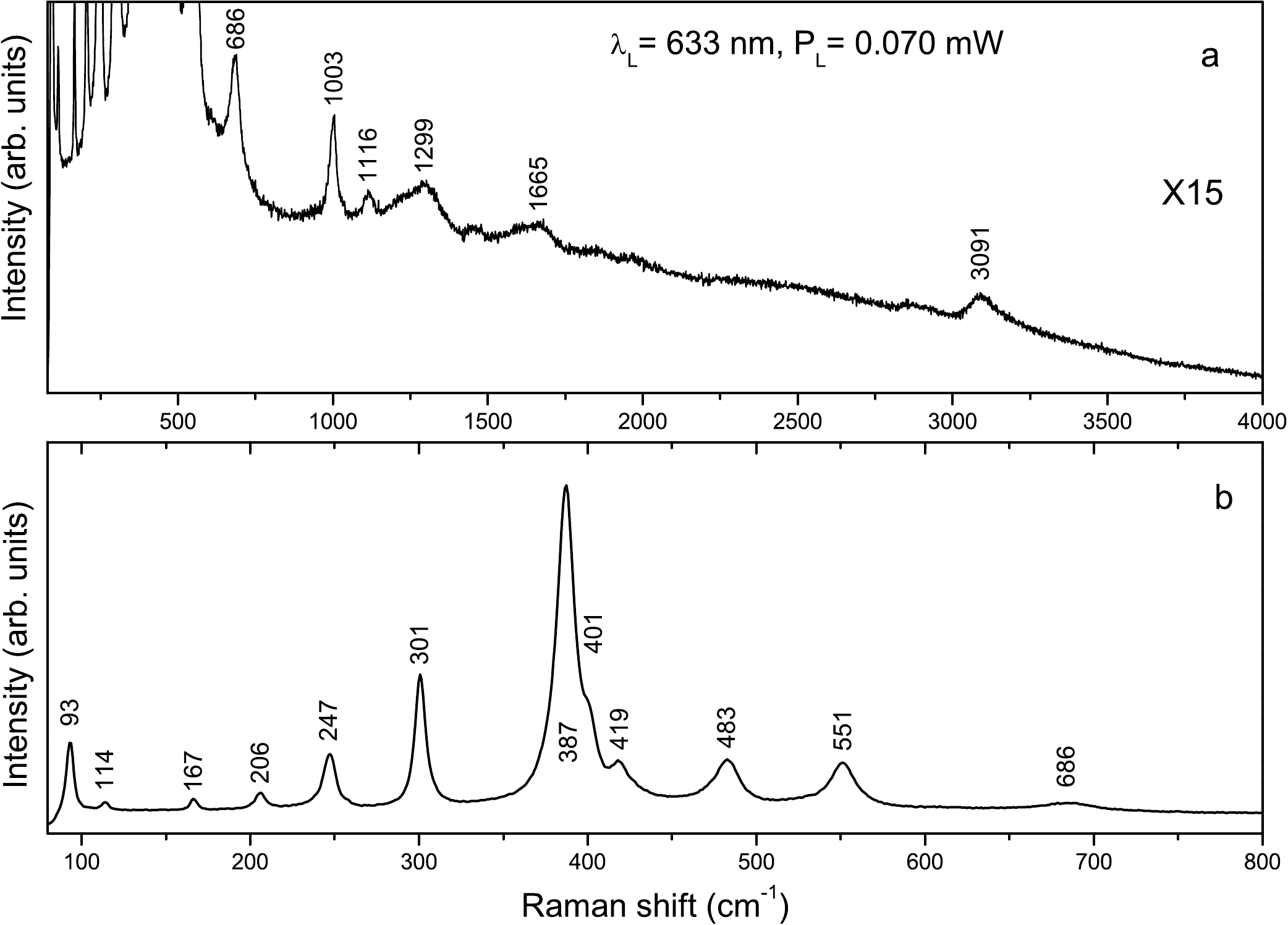}
\caption{\label{powder} (a) Raman spectrum of synthetic powder of goethite in the region $80$~cm$^{-1}$ -- $4000$~cm$^{-1}$; (b) The low-frequency part of the spectrum ($80$~cm$^{-1}$ -- $800$~cm$^{-1}$, the region where one-phonon Raman scattering by iron and oxygen vibrations is expected). The spectrum in (a) is multiplied by a factor of $15$ compared to the one in (b).}
\end{figure}

A spectrum obtained by synthetic powder of goethite (commercially available "Bayferrox910", Merck) is shown in Fig.~\ref{powder}. It coincides with the one published in Ref.~\onlinecite{Hanesch_2009}. The attempts to identify the symmetry of the lines comparing their relative intensity in the spectra obtained in parallel ($I_{\text{par}}$) and in crossed ($I_{\text{cr}}$) polarization (using the fact that the depolarization ratio for gas of molecules or randomly oriented particles $\rho = I_{\text{cr}}/I_{\text{par}}$ depends on the symmetry of the lines: $\rho = 3/4$ for B$_{1g}$, B$_{2g}$ and B$_{3g}$ lines and $0 \leq \rho \leq 3/4$ for A$_{g}$ lines) were unsuccessful as these two spectra were identical, showing that the scattered light is completely depolarized. This observation can be explained assuming that the penetration depth of the incident light is much larger than the submicronic size of the powder particles (their acicular shape and predominant size was determined by SEM as $0.1~\mu$m $\times~0.6~\mu$m).

\begin{figure}[htbp]
\centering
\includegraphics[width=\linewidth]{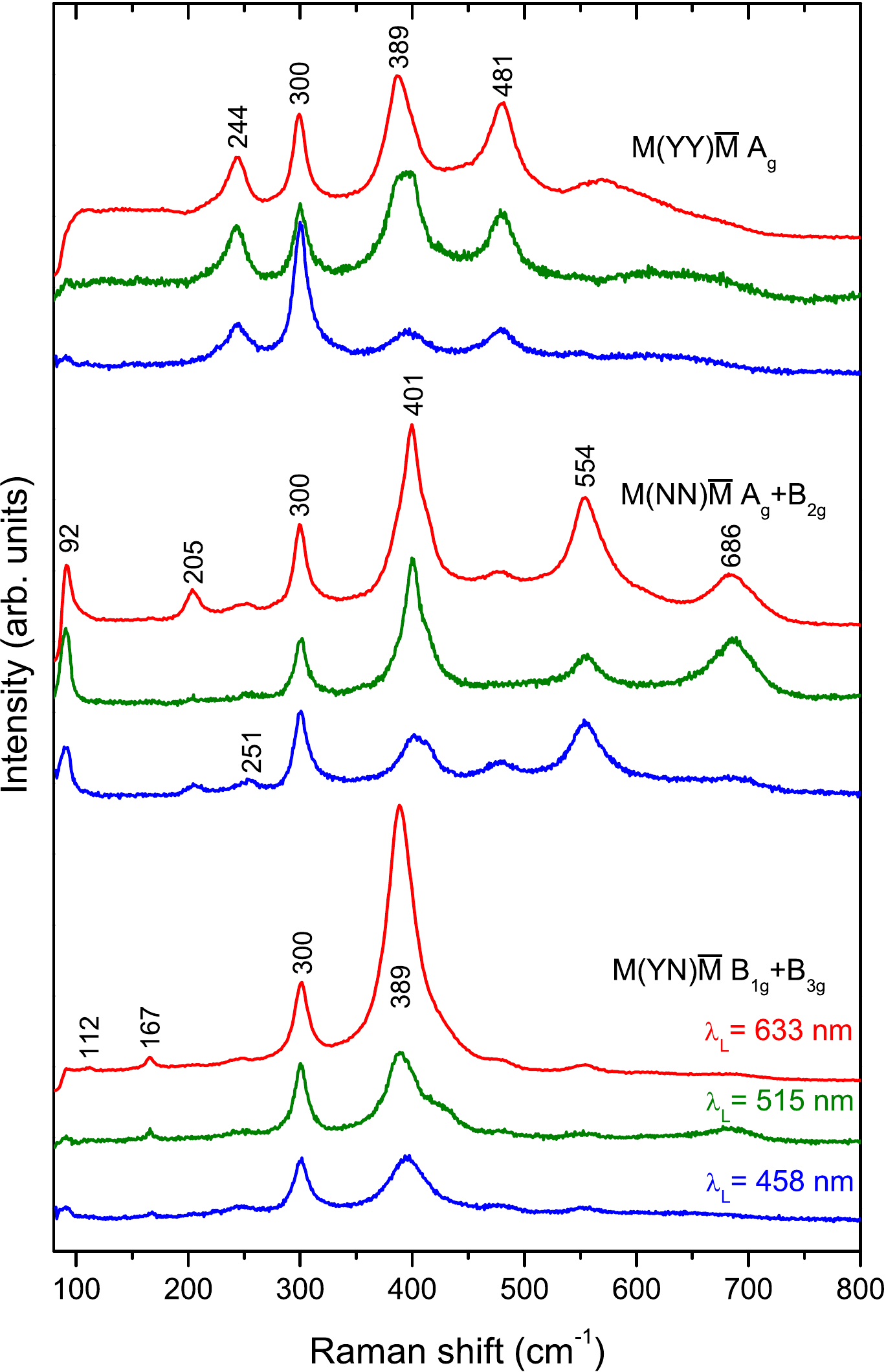}
\caption{\label{ores}Polarized Raman spectra of goethite ores. The wavelengths of the used laser excitation are shown in the figure. For each spectrum the scattering configuration (in Porto notations) and line symmetries, allowed for this configuration, are shown. $M$ and $N$ are two mutually perpendicular unknown directions in $(010)$ plane.}
\end{figure}

The polarized Raman spectra of goethite ores are shown in Fig.~\ref{ores}. As can be seen from electron microscope image (Fig.~\ref{SEM1}) in the scale of the laser spot the small needle-like microcrystals have nearly parallel long edges and this direction identifies the $[010]$ crystallographic direction ($y$-axis). However, the orientation of the other two directions, $[100]$ and $[001]$, for each microcrystal in the laboratory frame is probably arbitrary. This leads to the possibility that  only three qualitatively different polarized Raman spectra to be obtained: $M(YY)\bar{M}$, $M(NN)\bar{M}$, and $M(YN)\bar{M}$. The scattering configuration is described using the Porto notations. The first and the last symbol are the direction of the propagation of the incident and scattered light, whereas the symbols in brackets are the direction of the polarization of the incident and scattered light, respectively. Here $Y$ is the $[010]$ direction and $M$ and $N$ are two mutually perpendicular unknown directions within $(010)$ plane. The spectra obtained with different laser excitation are similar showing that the resonance effects are weak. From Fig.~\ref{ores} it is seen that the observed Raman lines can be sorted into three groups. The lines at $244$, $300$, $389$, and $481$~cm$^{-1}$ have $A_{g}$ symmetry, the ones at $92$, $205$, $300$, $401$, $554$, and $686$~cm$^{-1}$ have $A_{g}$ or $B_{2g}$ symmetry, and the ones at $112$, $167$, $300$, and $389$~cm$^{-1}$ have $B_{1g}$ or $B_{3g}$ symmetry. However, due to the very small size of the goethite crystals in the ore and their defect nature, the width of the lines is large making possible the observation only of the strongest lines. Also the recognition of the close positioned lines (e.g. the lines in the region near $400$~cm$^{-1}$) is difficult. The roughness of the surface also can contribute to deviations from the selection rules (expected for perfect crystal).

\begin{figure}[htbp]
\centering
\includegraphics[width=\linewidth]{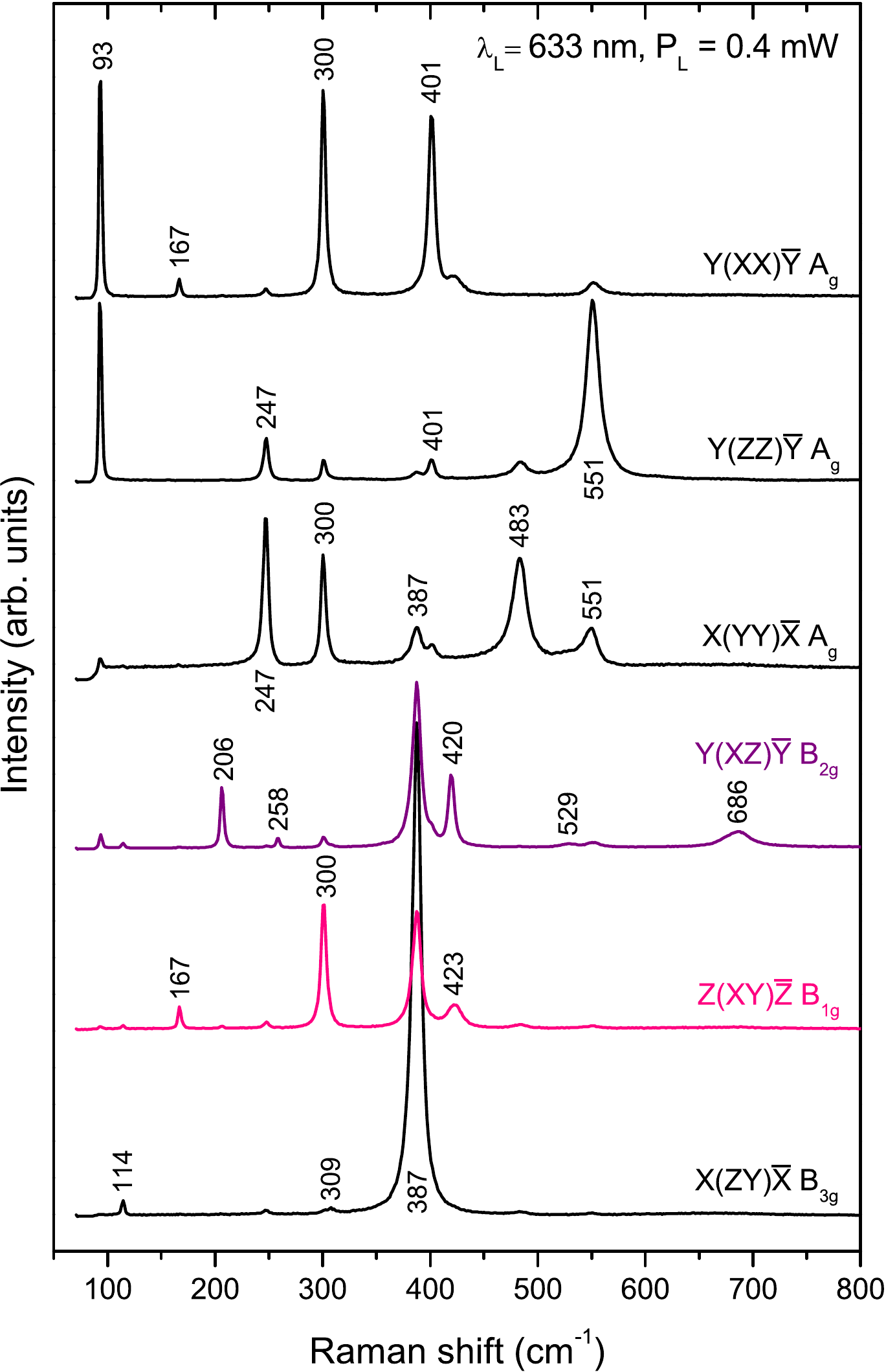}
\caption{\label{polarized}Polarized Raman spectra of goethite single crystals (low-frequency part). The wavelength of the laser excitation is $\lambda_{L} = 633$~nm. For each spectrum the scattering configuration (in Porto notations) and line symmetry, allowed for this configuration, are shown.}
\end{figure}

\begin{figure}[htbp]
\centering
\includegraphics[width=\linewidth]{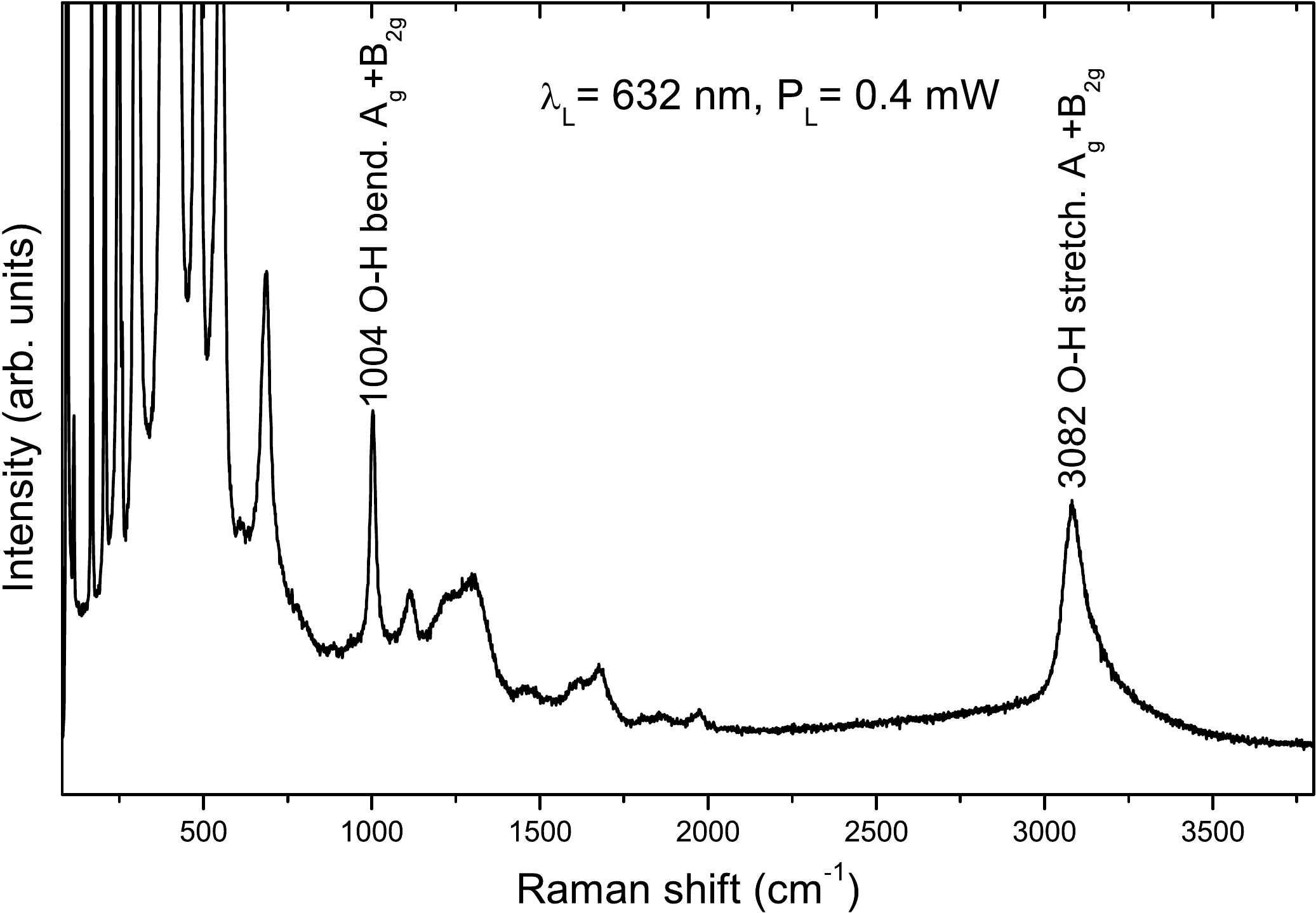}
\caption{\label{RamanOH}Non-polarized Raman spectrum obtained from a $(010)$ surface of goethite single crystal. The wavelength of the laser excitation is $\lambda_{L} = 633$~nm. The two Davydov pairs ($A_{g}$ and $B_{2g}$ modes) of hydrogen vibrations in the O-H groups are indicated.}
\end{figure}

The Raman spectra, obtained from single crystals of goethite, are shown in Figs.~\ref{polarized},~\ref{RamanOH}.

\begin{figure}[htbp]
\centering
\includegraphics[width=\linewidth]{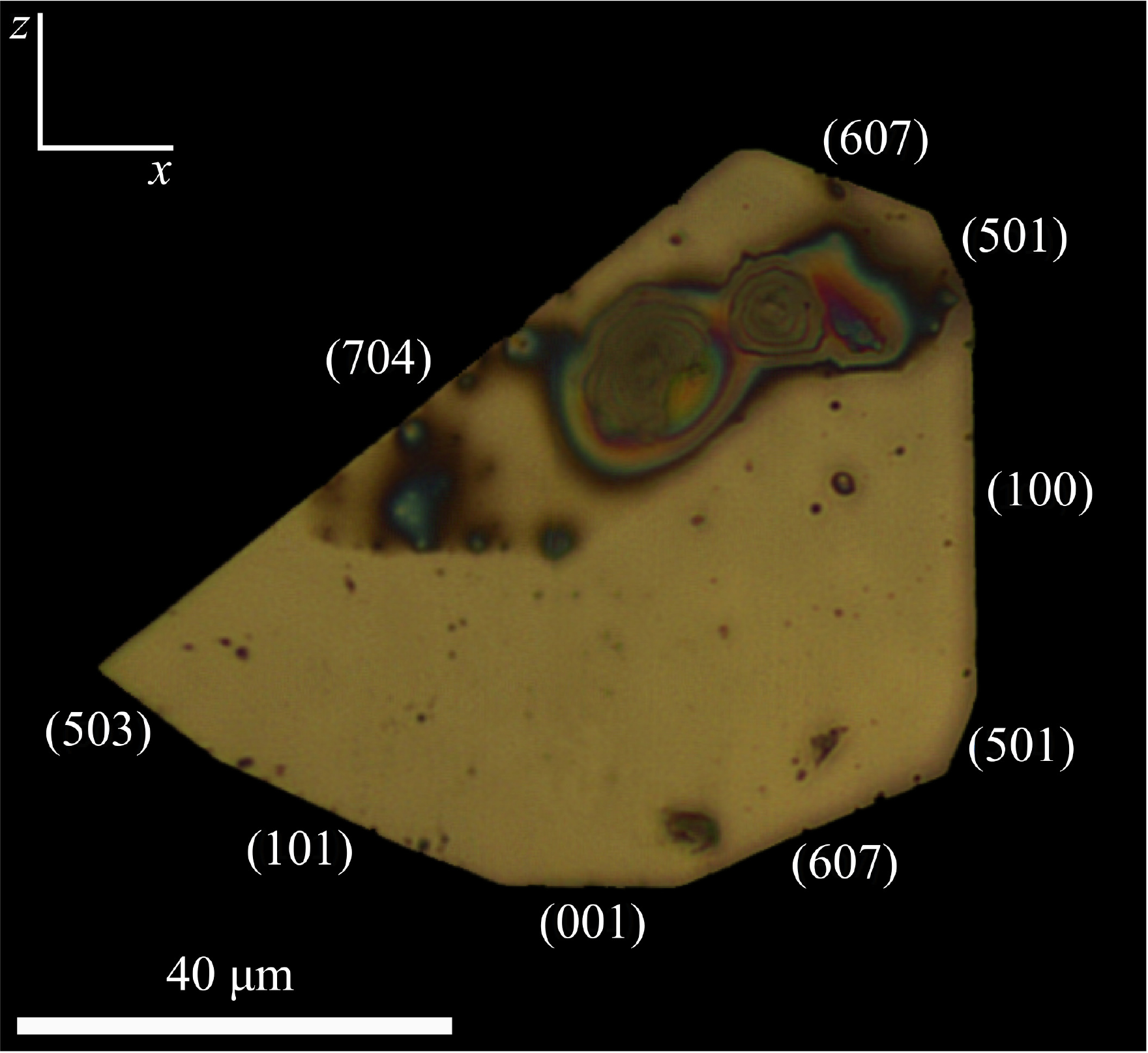}
\caption{\label{010face1} Optical photo of $(010)$ face of goethite crystal done with a $\times 50$ microscope objective. The Miller indices of the vertical planes, forming the edges of the $(010)$ face are indicated.}
\end{figure}

Before discussing the spectra in details, we will explain how the $[100]$ ($x$-axis) and $[001]$ ($z$-axis) directions were determined. In the case of orthorhombic crystal this is not trivial because the selection rules predict equal number of lines in $xy$ ($4B_{1g}$) and $zy$ ($4B_{3g}$) spectra. Also one and the same set of lines ($8A_{g}$) can be observed in $xx$ and $zz$ spectra. Therefore, the $x$ and $z$ directions are spectroscopically indistinguishable. The morphology of the needle-like crystals determines easily only the $y$ axis. The cross-section of the crystals (perpendicular to $y$ axis) looks rather round and it contains many edges along different crystallographic directions (not only of $\{101\}$ type that can be concluded from structural considerations, see Fig.~\ref{structure1}). Therefore we were looking for the presence of mutually perpendicular edges on $(010)$ faces ($xz$-planes) of different vertically aligned needle-like crystals. Then we obtained two spectra in parallel polarization (parallel and perpendicular to those edges) and one in crossed polarization. Thus, we succeeded in observing one set of lines (A$_{g}$ ones) in spectra in parallel ($xx$ and $zz$) polarization and another set of lines ($B_{2g}$) in crossed ($xz$) polarization (see Fig.~\ref{polarized}). After that from the photos of these faces we measured the angles between the $[100]$ and $[001]$ edges and the other edges. Knowing the lattice parameters, the angles for all $[h0l]$ edges can be calculated. Comparing the measured and calculated angles, the $x$ and $z$ directions as well all other edges (actually the ends of the vertical $(h0l)$ faces) can be identified (see Fig.~\ref{010face1}). After that, placing the needle-like crystals horizontally, we measured spectra in parallel polarization in direction perpendicular to the long edges as long as the measured spectrum coincide with one of the already known $xx$ and $zz$ spectra. Finally, we succeed in finding the $(100)$ and $(001)$ faces and measured the missing up to then $xy$ and $zy$ spectra, where only the lines with $B_{1g}$ and $B_{3g}$ symmetry can be observed.

On the basis of polarized spectra shown in Fig.~\ref{polarized}, the observed Raman lines can be easily discriminated by symmetry. The experimental frequencies, including those recorded on the synthetic powder and ores, are listed  in Table~\ref{tab1} and there they are compared to the LDC values obtained with the shell-model and \textit{ab initio} DFT.  The corresponding calculated amplitude vectors (from both calculating methods) for all Raman-active modes from the $\Gamma$-point of the Brillouin zone are given in the Supporting Information (SI) as xyz files, and could be visualized with the Jmol software. Evidently,  the proposed assignment of the experimental Raman features to specific mode symmetry corroborates to a good precision the theoretical expectations of the shell-model and DFT. It is also supported by the earlier GGA+$U$ calculations of Blanchard~{\it et~al.},\cite{Blanchard_2013} though atomic displacement vectors have not been reported there. Therefore, our further analysis will be based on the normal modes calculated in the present work.
\begin{table}[htbp]
\caption{\label{tab1}Comparison between the frequencies (in cm$^{-1}$) of the experimentally observed Raman lines (in the spectra of the synthetic powder, ores and the single crystals, respectively) and the calculated frequencies: shell model and DFT.}
\begin{ruledtabular}
\begin{tabular}{rcrrrrr}
&
&
\multicolumn{3}{c}{Experimental data} &
\multicolumn{2}{c}{LDC} \\
\cline{3-5}\cline{6-7}
& 
\multicolumn{1}{c}{Line}   &
\multicolumn{1}{c}{Synth.} & 
\multicolumn{1}{c}{Ores}   & 
\multicolumn{1}{c}{Single} & 
\multicolumn{1}{c}{Shell}  & 
\multicolumn{1}{c}{DFT}    \\

\multicolumn{1}{c}{No}.      &
\multicolumn{1}{c}{symmetry} &
\multicolumn{1}{c}{powder}   &
&
\multicolumn{1}{c}{cryst.}   &
\multicolumn{1}{c}{model}    & \\
\hline
 1  & $A_{g}(1)$  &    $93$ &   $92$ &   $93$ &   $90$ &  $123$ \\
 2  & $A_{g}(2)$  &   $247$ &  $244$ &  $247$ &  $237$ &  $261$\\
 3  & $A_{g}(3)$  &   $301$ &  $300$ &  $300$ &  $355$ &  $309$\\
 4  & $A_{g}(4)$  &   $401$ &  $401$ &  $401$ &  $389$ &  $426$ \\
 5  & $A_{g}(5)$  &   $483$ &  $481$ &  $483$ &  $418$ &  $490$\\
 6  & $A_{g}(6)$  &   $551$ &  $554$ &  $551$ &  $548$ &  $553$\\
 7  & $A_{g}(7)$  &  $1003$ & $1000$ & $1004$ & $1005$ & $1071$\\
 8  & $A_{g}(8)$  &  $3091$ & $3120$ & $3082$ & $3082$ & $3035$\\
 9  & $B_{1g}(1)$ &   $167$ &  $167$ &  $167$ &  $193$ &  $169$\\
10  & $B_{1g}(2)$ &         &  $300$ &  $300$ &  $303$ &  $305$\\
11  & $B_{1g}(3)$ &         &        &  $423$ &  $490$ &  $412$\\
12  & $B_{1g}(4)$ &         &        &        &  $744$ & $748$\\
13  & $B_{2g}(1)$ &   $206$ &  $205$ &  $206$ &  $222$ &  $210$\\
14  & $B_{2g}(2)$ &         &  $251$ &  $258$ &  $271$ &  $278$\\
15  & $B_{2g}(3)$ &         &        &  $356$ &  $370$ &  $351$\\
16  & $B_{2g}(4)$ &   $419$ &        &  $420$ &  $442$ &  $399$\\
17  & $B_{2g}(5)$ &         &        &  $529$ &  $467$ &  $574$\\
18  & $B_{2g}(6)$ &   $686$ &  $686$ &  $686$ &  $600$ &  $651$\\
19  & $B_{2g}(7)$ &  $1003$ & $1000$ & $1004$ & $1028$ & $1064$\\
20  & $B_{2g}(8)$ &  $3091$ & $3120$ & $3082$ & $3091$ & $3027$\\
21  & $B_{3g}(1)$ &   $114$ &  $112$ &  $114$ &  $123$ &  $97$\\
22  & $B_{3g}(2)$ &         &        &  $309$ &  $314$ &  $297$\\
23  & $B_{3g}(3)$ &   $387$ &  $389$ &  $387$ &  $463$ &  $422$\\
24  & $B_{3g}(4)$ &         &        &        &  $756$ & $743$\\
\end{tabular}
\end{ruledtabular}
\end{table}

\begin{figure}[htbp]
\centering
\includegraphics[width=\linewidth]{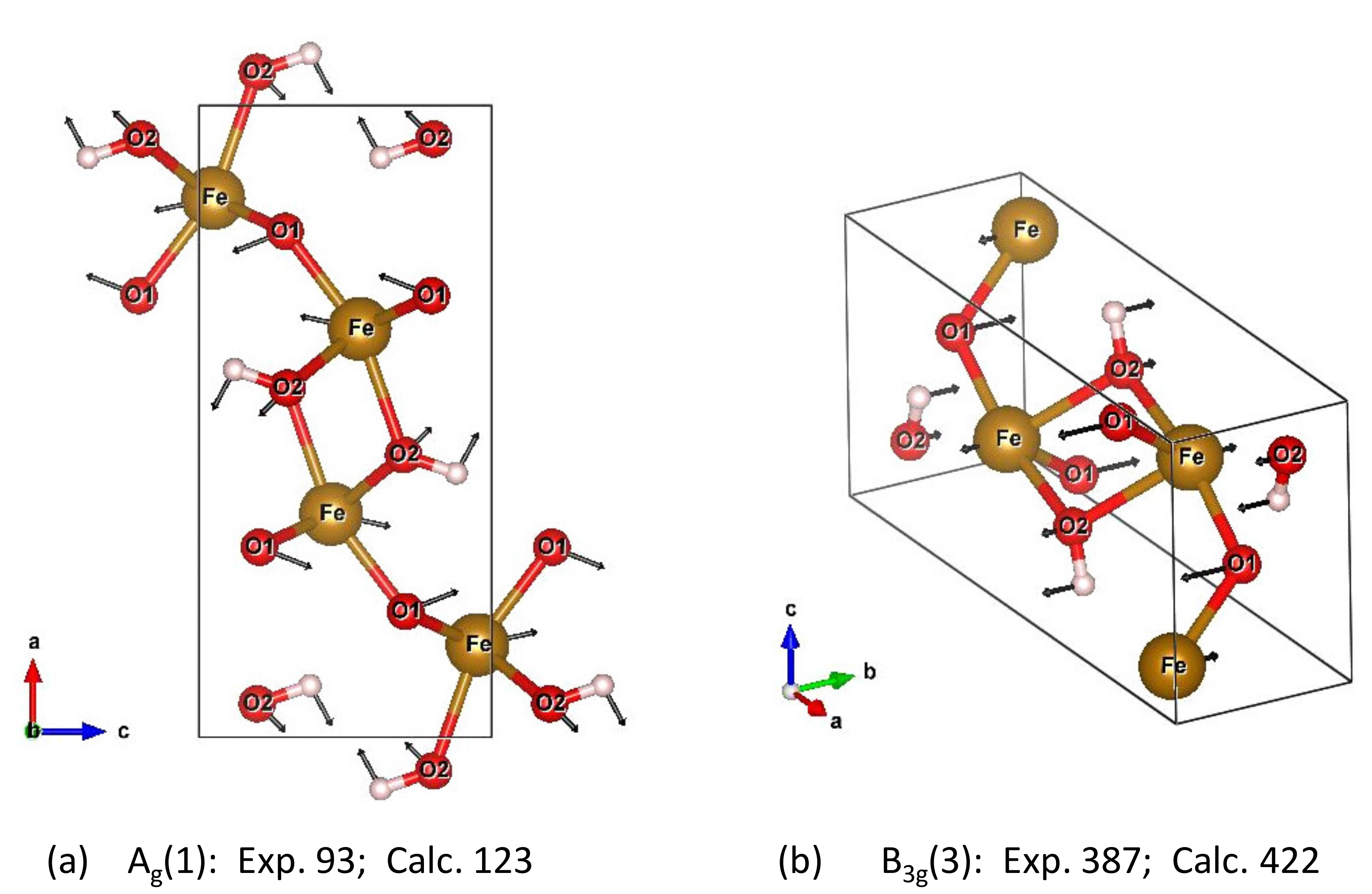}
\caption{\label{modes} Calculated atomic displacement vectors for the lowest-frequency $A_g(1)$ mode (a), and for the most intense $B_{3g}(3)$ mode (b). The experimental (Exp.) and calculated by DFT (Calc.) frequencies of the two vibrations are also given in the figure.}
\end{figure}

The presence of a Raman line below $100$~cm$^{-1}$ (the $A_{g}(1)$ mode at $93$~cm$^{-1}$) is not typical for crystal structures, in which the heaviest atom is iron. As shown in Fig.~\ref{modes}(a) this mode corresponds to a libration, i.e. a solid rotation of the double chains about $y$-axis. The displacement pattern of most of the Raman-active vibrations is complex and will not be discussed in details. Instead, we will focus  on the strongest feature in the spectra, the $B_{3g}(3)$ mode at $387$~cm$^{-1}$. According to the shell-model and DFT calculations, this mode is dominated by an O1 vibration along $y$, as depicted in Fig.~\ref{modes}(b). Due to the specific atomic arrangement in $\alpha$-FeOOH the O1 atoms move along the bisector of the Fe-O1-Fe angle ($\approx 124^{\circ}$). Therefore, this vibration could be characterized as a mixture of Fe-O1-Fe bond angle bending and a symmetric Fe-O1 stretch, since both iron-oxygen bonds in the Fe-O1-Fe linkage are modulated in-phase. 

In Fig.~\ref{RamanOH} the lines observed at $1004$~cm$^{-1}$ and $3082$~cm$^{-1}$ correspond to the two Davydov pairs of $A_{g}+B_{2g}$ modes of hydrogen vibrations. The other features with irregular shape between $1000$~cm$^{-1}$ and $2000$~cm$^{-1}$ originate from two- (and three-) phonon scattering.

\begin{figure}[htbp]
\centering
\includegraphics[width=\linewidth]{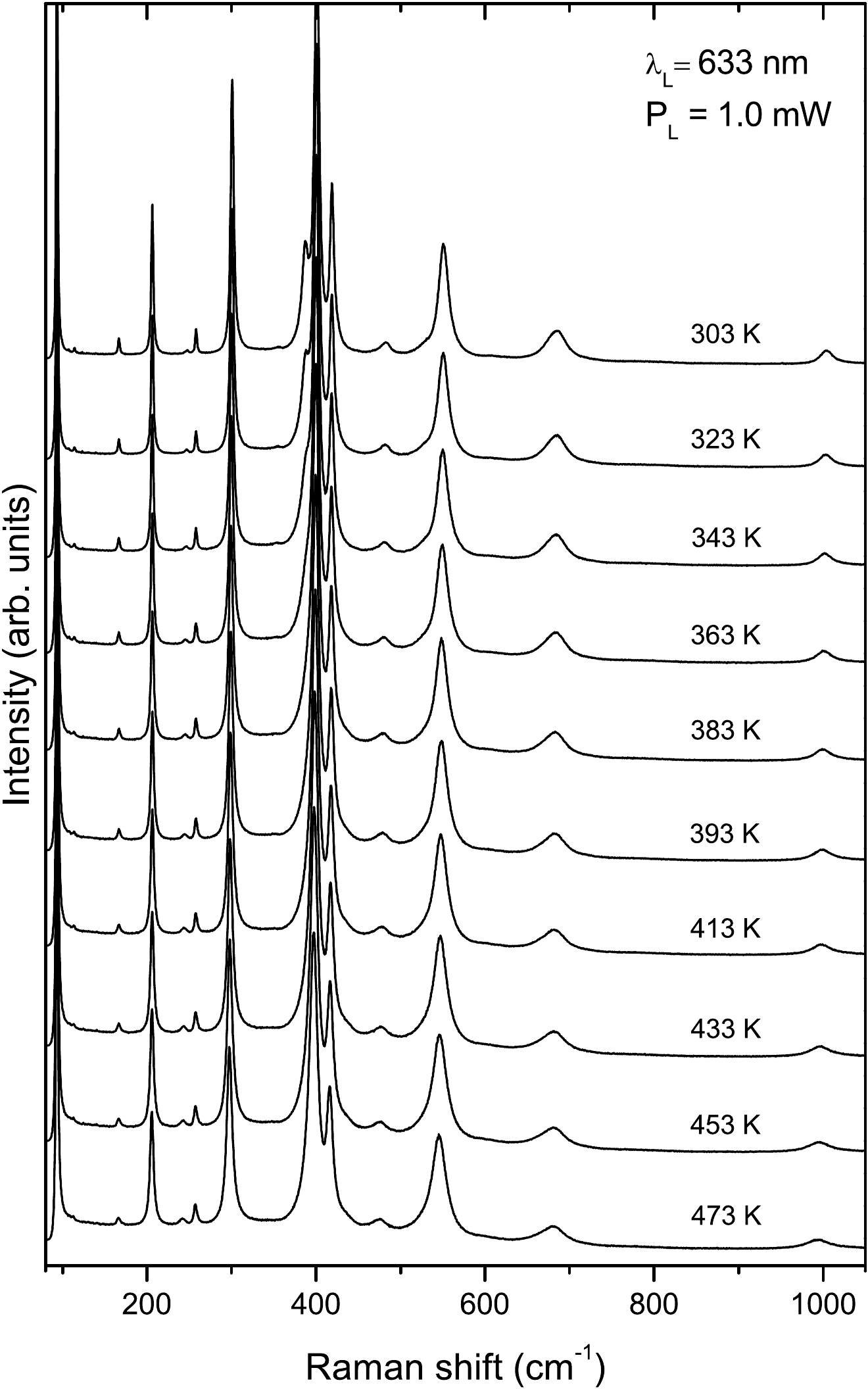}
\caption{\label{Tdepblackup}Raman spectra obtained at different temperatures from $(101)$ face of a goethite crystal and polarization of the incident laser light perpendicular to $[010]$ direction. No analyzer. In this scattering configuration the lines with $A_{g}$ and $B_{2g}$ symmetry dominate in the spectra.}
\end{figure}

\begin{figure}[htbp]
\centering
\includegraphics[width=\linewidth]{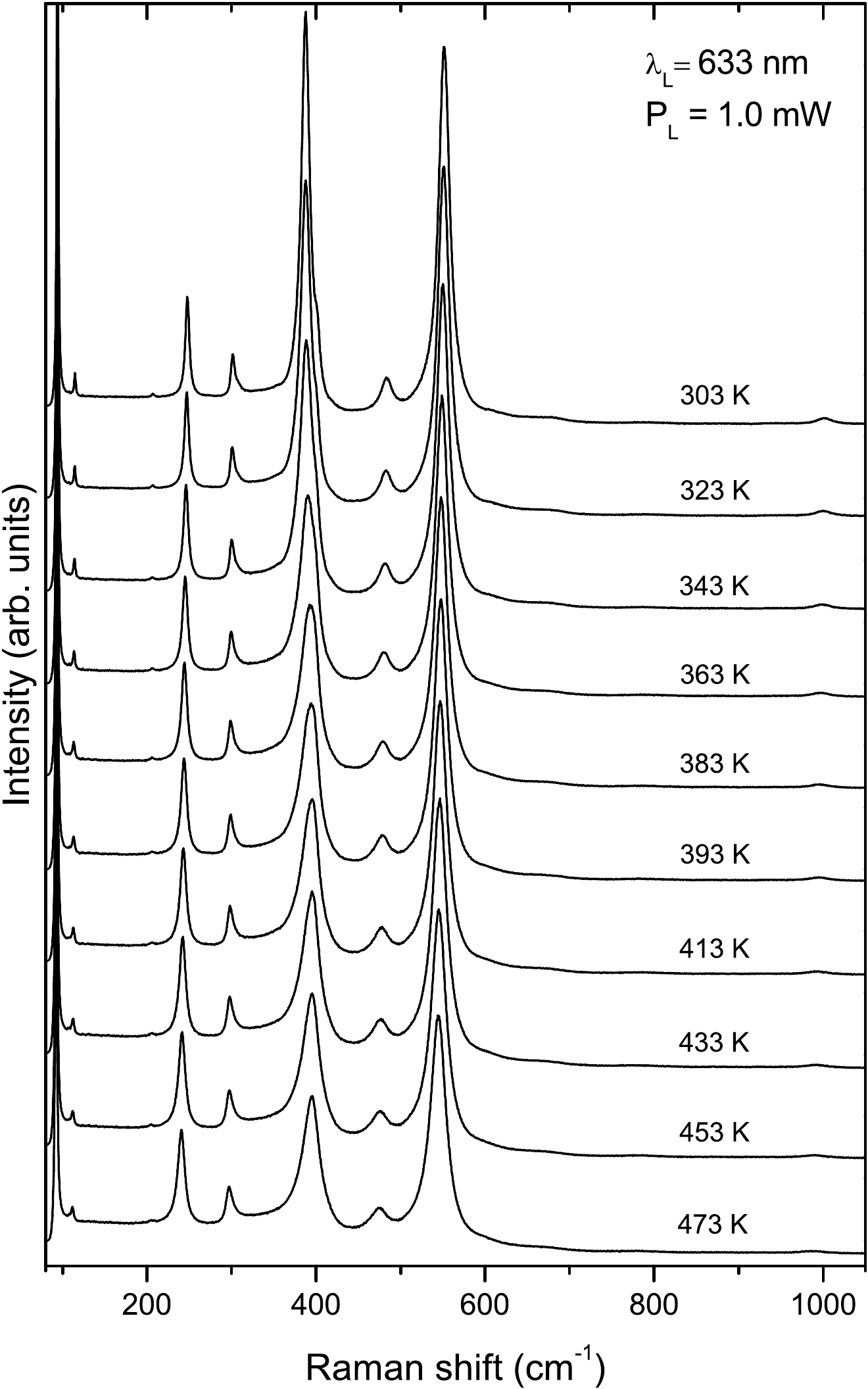}
\caption{\label{Tdepwhitedown}Raman spectra obtained at different temperatures from $(100)$ face of a goethite crystal and polarization of the incident laser light perpendicular to $[010]$ direction. No analyzer. In this scattering configuration the lines with $A_{g}$ and $B_{3g}$ symmetry are visible.}
\end{figure}

In Figs.~\ref{Tdepblackup},~\ref{Tdepwhitedown} are presented Raman spectra obtained at different temperatures between $303~$K and $473~$K. They were obtained from lying needle-like crystals with polarization of the laser light perpendicular to their long edge and without an analyzer.

\begin{figure}[htbp]
\centering
\includegraphics[width=\linewidth]{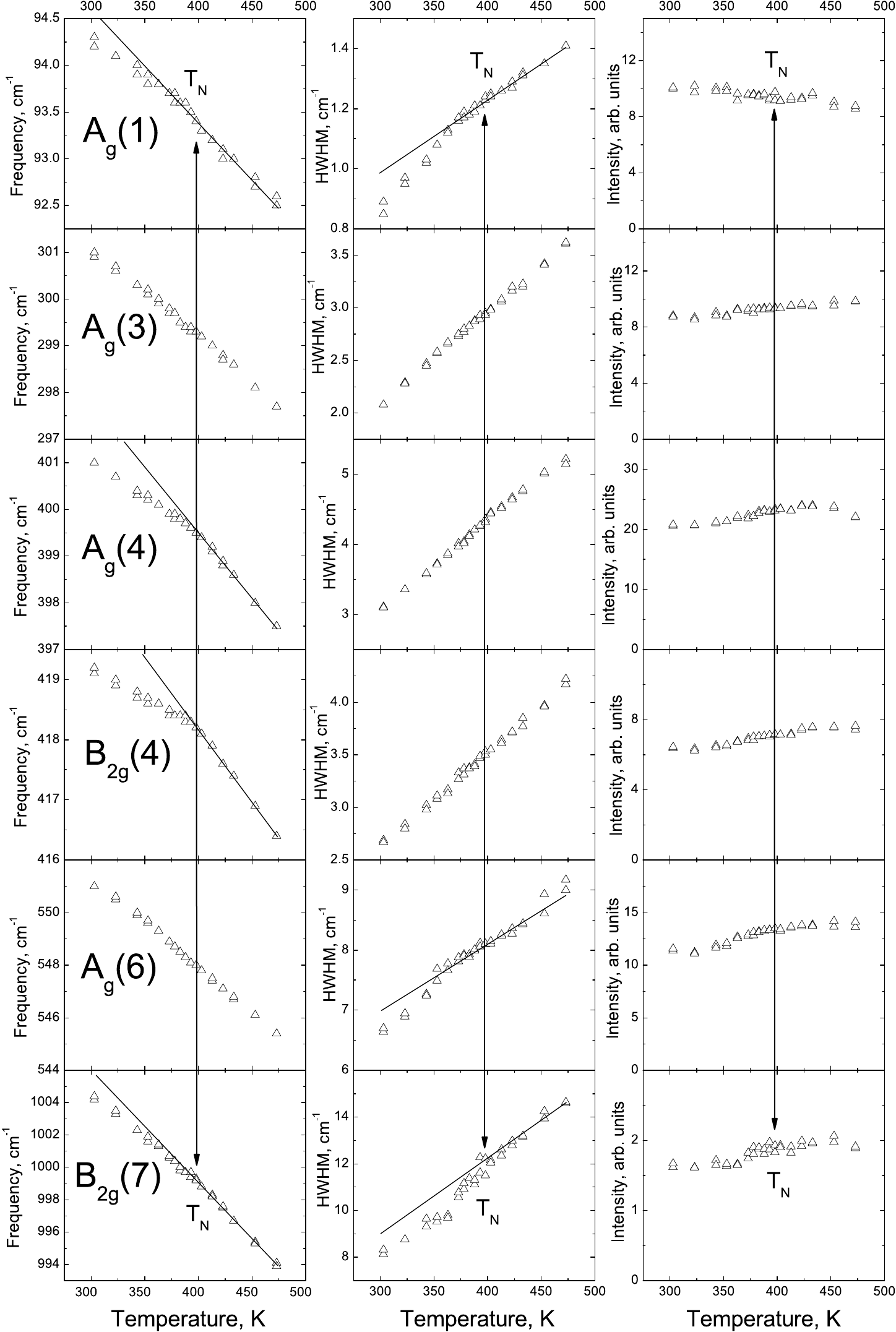}
\caption{\label{Tdep6lines} Temperature dependencies of the parameters (position, HWHM (half-width at half maximum) and integral intensity) of some of the lines. Their assignment is given in Table~\ref{tab1}. For the $A_{g}(1)$ line the parameters were calculated from the spectra, given in Fig.~\ref{Tdepwhitedown}. For the other five lines the parameters were calculated from the spectra, given in Fig.~\ref{Tdepblackup}. The two symbols at each temperature correspond to values of the fits from two different spectra. Some of the symbols coincide. The lines tracing some dependencies are just guide to the eye.}
\end{figure}

Analyzing the relative intensities of the observed lines it can be concluded that the spectra presented in Fig.~\ref{Tdepblackup} are obtained from $(101)$ face ($A_{g}$ and $B_{2g}$ lines dominate in the spectra). These spectra were obtained on heating. The spectra presented in Fig.~\ref{Tdepwhitedown} were obtained from a $(100)$ face ($A_{g}$ and $B_{3g}$ lines dominate in the spectra). These latter spectra were obtained on cooling. No any sign of chemical decomposition during the heating/cooling cycle and the spectra obtained at room temperature before heating and after cooling were identical. The most intense lines in the spectra were fitted with Lorentzians. The temperature dependencies of lineshape parameters (position, half-width at half-maximum -- HWHM, and integrated intensity) for some of these modes are shown in Fig.~\ref{Tdep6lines}. It is clear that for $A_{g}(1)$, $A_{g}(4)$, $B_{2g}(4)$, and $B_{2g}(7)$ modes the slope of the frequency-temperature curve is discontinuous at $T_{\text{N}}$. Likewise, a discontinuity is found at $T_{\text{N}}$ in the slope of HWHM-temperature dependence for the $A_{g}(1)$, $A_{g}(6)$, and $B_{2g}(7)$ modes. These ``anomalies'' infer for a considerable spin-lattice coupling in $\alpha-$FeOOH.  The spectral changes near $T_{\text{N}}$, however, are better pronounced and more informative for the most intense $B_{3g}(3)$ mode at $387$~cm$^{-1}$.

\begin{figure}[htbp]
\centering
\includegraphics[width=\linewidth]{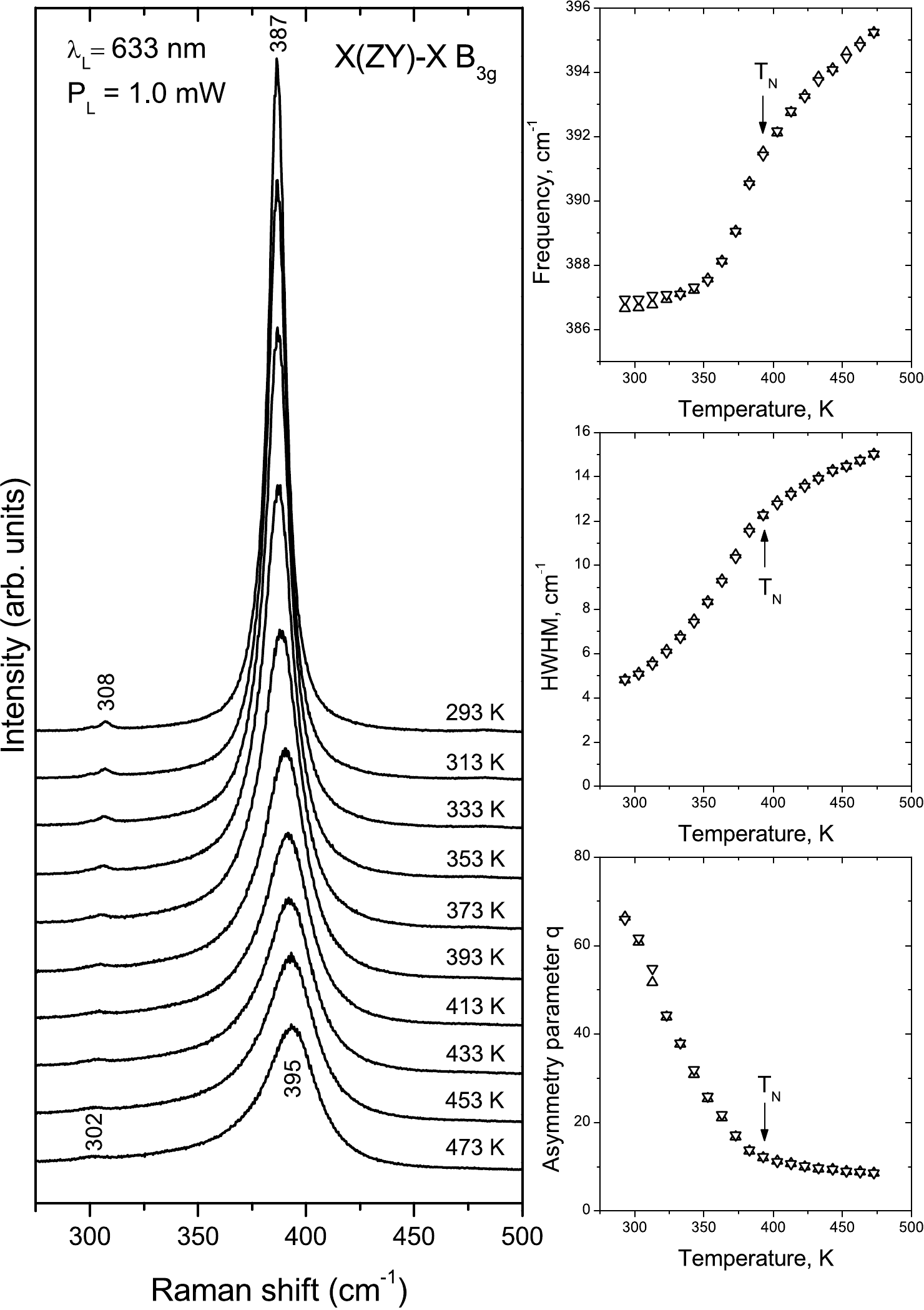}
\caption{\label{Tdep387} (left panel)---Raman spectra obtained from goethite single crystal at different temperatures in $X(ZY)\bar{X}$ scattering configuration (only lines with $B_{3g}$ symmetry are allowed); (right panel)---The temperature dependence of the parameters (position, HWHM, and asymmetry parameter) calculated after a fit of the $B_{3g}(3)$ line with Fano profile. Up-pointing triangle symbols correspond to the spectra obtained on heating, down-pointing triangle symbols correspond to the spectra obtained on cooling.}
\end{figure}

The  fit of the  $B_{3g}(3)$ line with Lorentzian shape is unsatisfactory, with a strong and asymmetric spectral residuum. In order to obtain a relevant description of the $B_{3g}(3)$ lineshape, we performed additional measurements of its temperature evolution  (both on heating and cooling) in $X(ZY)\bar{X}$ scattering configuration, in a narrower spectral window by using higher resolution diffraction grating. In this configuration only $B_{3g}$ symmetry is allowed, so the possible superimposition with other closely positioned lines (such as $A_{g}(4)$ at $401$~cm$^{-1}$ and $B_{2g}(4)$ at $420$~cm$^{-1}$) is excluded, and thus we were able to record a pure signal from the $B_{3g}(3)$ mode only. The corresponding spectra are shown in the left panel of Fig.~\ref{Tdep387}.

It is seen that with the increase of the temperature from $293~$K to $473~$K, the frequency of the weak $B_{3g}(2)$ line decreases from $308$ to $302$~cm$^{-1}$, whereas the frequency of the $B_{3g}(3)$ mode increases from $387$ to $395$~cm$^{-1}$. The asymmetric shape of the $B_{3g}(3)$ line is apparent in the spectra at high temperatures. Therefore, we fitted it with an asymmetric Fano profile, $\displaystyle I( \nu )= I_{0}\frac{\left(1+ \frac{ \nu - \nu_{0} }{q \Gamma } \right)^2}{1+ \left(\frac{\nu - \nu_{0}}{\Gamma}\right)^2 }$, where the $q$ is the asymmetry parameter. When $q \rightarrow \infty$, the Fano profile converts to the Lorentzian shape, as the $\nu_{0}$, $\Gamma$, and $I_{0}$ are the position, HWHM, and peak intensity of the Lorentzian, respectively. The calculated parameters from the fit are given in the right panel in Fig.~\ref{Tdep387}. Evidently, the spectral shape parameters display considerable changes around $T_{\text{N}}$. The integrated intensity of the line (not shown in the figure) also displays an abrupt increase below $T_{\rm N}$, which is similar to the temperature behavior observed for specific Raman lines in the magnetoelectric compound Cu$_2$OSeO$_3$.\cite{gnezdilov2010} In the later case, the intensity anomaly has been interpretted as an increase of the dynamic electric polarizability of the material due to a contribution from the magnetoelectric effect. Therefore, it is tentative to interpret the high intensity of the $B_{3g}(3)$ line below $T_{\rm N}$ as a signature for a magnetoelectric succeptability of $\alpha$-FeOOH in its magnetically ordered phase, in agreement with the theoretical predicitons of Ref.~\onlinecite{Ter_Oganessian_2017}. 

Especially interesting is the temperature evolution of the asymmetry parameter $q$ ($q<0$) of the $B_{3g}(3)$ line. With the increase of temperature, $q$ decreases in modulus, and changes weakly above $T_{\text{N}}$. This results in a more pronounced Fano-shape of $B_{3g}(3)$ line above $T_{\text{N}}$, which indicates to the presence of an excitation continuum whose spectral density undergoes substantial redistribution at the AF/PM transition. It is evident from Fig.~\ref{RamanOH} that the low-frequency phonons overlay a wide scattering band, extending up to  $\approx 1000$~cm$^{-1}$, with aparent maximum at $\approx 400$~cm$^{-1}$ -- in a close proximity to the spectral position of the $B_{3g}(3)$ mode. It is tentative, therefore, to assume that the Fano-shape of the $B_{3g}(3)$ line is a result of interaction between that mode and the excitations from the underlaying background. The origin of this scattering continuum, however, is ellusive since systematic studies of the electronic and magnetic excitations in $\alpha-$FeOOH are still lacking. Nevertheless, two plausible hypotheses for the asymmetric shape of the $B_{3g}(3)$ mode could be inferred from the existing works on structure, transport properties and magnetic ordering in goethite. 

First, and more coherent scenario is based upon coupling of the $B_{3g}(3)$ phonon with magnetic exchange excitations. Ter Oganessian {\it et~al.}\cite{Ter_Oganessian_2017} have reported extensive GGA+$U$ calculations of the exchange interactions in $\alpha-$FeOOH and have established that the strongest exchange, $J_2 = 48.1$~meV$ = 388$~cm$^{-1}$, corresponds to the Fe-O1-Fe bridges connecting pairs of neighboring double chains. The exchange energy of a Fe-O1-Fe dimer is given by: $H_{\rm s-s}=J_2 \vec S_i \cdot \vec S_j = \frac12 J_2 (K(K+1)-2S(S+1))$, where  $\vec S_i$ and $\vec S_j$ are the spins of the $i$-th and $j$-th Fe atoms respectively, $S$ is the spin per Fe atom and K is the total spin of the dimer. Obviously, if the dimer was isolated, its ground state would corresponds to a singlet ($K=0$) and the first excited state -- to a triplet ($K=1$). Notably the energy of the singles-triplet transition $\Delta E = J_2$ matches precisely the spectral maximum of the background. Moreover, the singlet-triplet transition of the dimer is Raman-active through the Fleury-Elliot mechanism of exchange-assisted light scattering (usually referred to as ``two-magnon'' scattering).\cite{Fleury1968}  Since different Fe-O1-Fe dimers are not isolated, but interact between each other via exchange interactions of comparable magnitude, the singlet-triplet scattering would be manifested as a smeared band of a spectral width comparable to $J_2$, instead of a  discrete line. Therefore, it is tentative to assign the scattering background to magnetic excitations involving singlet-triplet excitations of the Fe-O1-Fe dimers.  Since $K=1$ spin state transforms as an axial vector, this kind of scattering is active in $B_{1g}+B_{2g}+B_{3g}$ irreducible representations of the $Pnma$ space group, and could be coupled to phonons of the corresponding symmetries, provided physical mechanisms of such a coupling are present.  As discussed above, the O1 displacement in the $B_{3g}(3)$ vibration modulates the Fe-O1-Fe bond-angle, as well as the Fe-O1 bond lengths (see Fig.~\ref{modes}(b)). Therefore, a linear spin-phonon coupling of the form $H_{\rm s-ph}=\vec u \cdot \nabla_uJ_2 \vec S_i \cdot \vec S_j$ is allowed for the $B_{3g}(3)$ mode, where $\vec u$ is the O1 displacement vector, $\nabla_uJ_2$ is the gradient of $J_2$ with respect to O1 displacement. Correspondingly, this spin-phonon interaction is likely to result in an asymmetric Fano-shape of the $B_{3g}(3)$ line, since the phonon is directly coupled with the spin excitation of the dimer. 

It is worthy to note that the mechanism of spin-phonon coupling of the $B_{3g}(3)$ mode in $\alpha-$FeOOH is qualitatively different from the spin-phonon interaction studied previously in perovskite manganites.\cite{Granado1998,Granado1999,Laverdiere2006} Phonons, displaying significant frequency shift below $T_{\rm N}$ in manganites, are associated with asymmetric Mn-O-Mn stretching, for which the linear spin-phonon coupling is forbidden by symmetry. Instead, these phonons are involved in a second-order spin-phonon interaction, given by $H_{\rm s-p} = \frac12 J'' M^2 u^2$, where $J''$ is the second derivative of the exchange integral with respect to the oxygen displacement $u$, and $M$ is the sublattice magnetization below $T_{\rm N}$. Thus, the second-order interaction is equivalent to an additional force-constant, which results in a frequency shifts  below $T_{\rm N}$, but does not affect the symmetric shape of the phonon line. The second-order spin-phonon coupling could be operative for the $B_{3g}(3)$ phonon in $\alpha-$FeOOH, as suggested from the pronounced frequency softening of this mode (see Fig.~\ref{Tdep387}). Due to the presence of a first-order interaction, however, the frequency shift of this vibration below  $T_{\rm N}$ depends not only on $J''$, like in manganites, but also on $J'^2/J$. Therefore, the first and the second derivatives of the exchange integral with respect to the O1 displacement could not be extracted independently from the measured values of the frequency shift.

In order to qualify the magnetic mechanism of $B_{3g}(3)$ line asymmetry, one should also answer why asymmetry of the phonon line is more prononced (i.e. the Fano $q$-parameter decreases in modulus) above the magnetic transition temperature. Recently, an extensive Raman study of the spin excitations in the antiferromagnetic insulator Cu$_2$OSeO$_3$ has been reported by Versteeg {\it et~al.}\cite{Versteeg2019} This compound consists of structurally isolated Cu$_4$ spin units, which give rise to a multitude of intra-cluster spin excitations, which persist well above the N\'eel temperature $T_{\rm N}$, when the antiferromagnetic correlations between different clusters is lost. In the case of $\alpha-$FeOOH the Fe-O1-Fe dimers are not isolated to such an extent as the Cu$_4$ clusters in Cu$_2$OSeO$_3$. Nevertheless, above $T_{\rm N}$ the antiferromagnetic correlations between different Fe-O1-Fe bridges are lost and the excitation spectral density will be shifted closer to the energy of singlet-triplet excitation of an isolated dimer and resspectively the $B_{3g}(3)$ phonon line.  Being superimposed on a background of a larger spectral density above $T_{\rm N}$, the $B_{3g}(3)$ phonon acquires more pronounced line asymmetry in PM phase compared to AF phase.

The second possible mechanism of the $B_{3g}(3)$ line asymmetry -- still very qualitative -- is the interaction of this phonon with thermally-activated charge carriers. $\alpha-$FeOOH  is  a charge-transfer insulator with a band gap of 2.5~eV\cite{Sherman2005} whose electric conduction is related to a thermally-activated hopping of small polarons.\cite{Porter_2018}  The O1 atoms mediate not only the exchange interaction but also the charge transport between Fe atoms, and the $B_{3g}(3)$ vibration could be effectively coupled to the charge carrier hopping. Below $T_{\rm N}$ the charge transport between the antiferromagnetically ordered Fe spins is largely suppressed due to the Hund's repulsion between electrons of opposite spins. Above transition temperature, however, the antiparallel spin arrangement is lost and correspondingly the carrier mobility increases, i.e. an  insulator-metal transition (coinciding with the AF-PM transition) cannot be excluded. This mechanism resembles the double-exchange model describing the charge conduction near the magnetic phase transition in doped LaMnO$_3$. The qualification of this scenario, however, requires additional experiments, like magnetotransport measurements near the transition temperature, and theoretical understanding of the excitation spectrum associated with the charge carriers in $\alpha-$FeOOH. 

\section{Conclusions}
Raman spectra (both non-polarized and polarized) were obtained on different samples of $\alpha$-FeOOH (goethite)  -- synthetic powder, ores and mineral single crystals. The symmetry representation of the observed Raman features was determined on the basis of polarization selection rules. The spectral lines were assigned to definite atomic vibrations by comparison with lattice-dynamical calculations, and 22 out of the 24 Raman-active modes were identified. The measurements of the Raman spectra in the temperature interval 293~K -- 473~K (including the temperature of the antiferromagnetic-paramegnetic transition at $T_{\text{N}} = 393$~K) reveals anomalous temperature behavior of the lineshape parameters of specific phonons around $T_{\text{N}}$, which evidences for a significant interaction between spin and lattice degrees of freedom. In particular, the $B_{3g}(3)$ mode at 387~cm$^{-1}$ reveals  strongly assymetric Fano-shape above $T_{\text{N}}$. This finding is most likely a signature of a strong coupling between this phonon and a continuum of magnetic excitations, whose spectral density is substantially redistributed above the transition temperature.  However, an interaction between the $B_{3g}(3)$ vibration with the thermally-activated charge carriers could also play role, and infers for a possible conductivity enhancement accompanying the magnetic transition. 

\begin{acknowledgments}
MVA, VGI, and NDT thank the support by the Bulgarian Ministry of Education and Science under contract D01-284/2019 (INFRAMAT) and by the European Regional Development Fund within the Operational Programme "Science and Education for Smart Growth 2014 -- 2020" under the Project CoE "National center of mechatronics and clean technologies" BG05M2OP001-1.001-0008-C01. MVA thanks the Alexander von Humboldt Foundation, Bonn (Germany) for the research fellowship, ensuring his stay at Freie Universitat Berlin. This work was partially supported by the bilateral Bulgarian-Russian project KP-06-15 funded by the Ministry of Education and Science, as well as received funding from the European Research Council (ERC) under the European Union’s Horizon 2020 research and innovation programme (grant agreement No. 819623). The authors thank Rositsa Titorenkova (Institute of Mineralogy and Crystallography-BAS, Bulgaria) and Jordan Kortenski (University of Mining and Geology, Bulgaria) for the supply of the ore samples. The helpful discussions with Milko N. Iliev are highly appreciated as well as the help from Anna Esther and Marti Gich with the electron microscopy work.
\end{acknowledgments}

\vspace*{-\baselineskip}
\section*{References}
\vspace*{-\baselineskip}

\end{document}